\documentclass[final,3p,times,twocolumn]{elsarticle}

\usepackage{lineno,hyperref}
\modulolinenumbers[5]
\usepackage{comment}
\usepackage{bm}
\usepackage{here}
\usepackage{color}
\journal{Journal of \LaTeX\ Templates}

\newcommand\red{\textcolor{black}}
\begin{document}

\begin{frontmatter}
\title{
Development of the detector simulation framework for the Wideband Hybrid X-ray Imager onboard FORCE
}
\author[ut]{Hiromasa Suzuki\corref{ca}}
\cortext[ca]{Corresponding Author}
\ead{suzuki@juno.phys.s.u-tokyo.ac.jp}

\author[ut]{Tsubasa Tamba}
\author[ut,re]{Hirokazu Odaka}
\author[ut,re]{Aya Bamba}
\author[tus]{Kouichi Hagino}
\author[um]{Ayaki Takeda}
\author[um]{Koji Mori}
\author[um]{Takahiro Hida}
\author[um]{Masataka Yukumoto}
\author[um]{Yusuke Nishioka}
\author[ku]{Takeshi G. Tsuru}

\address[ut]{Department of Physics, The University of Tokyo, Tokyo 113-0033, Japan}
\address[re]{Research Center for the Early Universe, The University of Tokyo, Tokyo 113-0033, Japan}
\address[tus]{Department of Physics, Tokyo University of Science, Noda, Chiba 278-8510, Japan}
\address[um]{Department of Applied Physics and Electronic Engineering, University of Miyazaki, Miyazaki 889-2192, Japan}
\address[ku]{Department of Physics, Kyoto University, Kyoto 606-8502, Japan}

\begin{keyword}
Astronomy, X-ray, detector response, Monte Carlo simulation, Silicon-on-insulator technology
\end{keyword}

\begin{abstract}

FORCE is a Japan-US space-based astronomy mission for an X-ray imaging spectroscopy in an energy range of 1--80 keV. The Wideband Hybrid X-ray Imager (WHXI), which is the main focal plane detector, will use a hybrid semiconductor imager stack composed of silicon and cadmium telluride (CdTe). \red{The silicon imager will be a certain type of the silicon-on-insulator (SOI) pixel sensor, named the X-ray pixel (XRPIX) series.} Since the sensor has a small pixel size (30--36 $\mu$m) and a thick sensitive region (300--500 $\mu$m), understanding the detector response is not trivial and is important \red{in order to optimize the camera design and to evaluate the scientific capabilities.}
We have developed a framework to simulate observations of celestial sources with semiconductor sensors.
Our simulation framework was tested and validated by comparing our simulation results to laboratory measurements using the XRPIX 6H sensor. The simulator well reproduced the measurement results with reasonable physical parameters of the sensor including an electric field structure, a Coulomb repulsion effect on the carrier diffusion, and arrangement of the degraded regions.
This framework is also applicable to future XRPIX updates including the one which will be part of the WHXI, as well as various types of semiconductor sensors.

\end{abstract}
\end{frontmatter}

%%%%%%%%%%%%%%%%%%%%%%%%%%%%%%%%%%%%%%%%55
\section{Introduction}
In X-ray astrophysics, we have to observe celestial sources with in-orbit observatories because of atmospheric absorption.
In order to obtain the emission spectrum of a celestial source correctly and to derive \red{its} physical parameters, it is essential to understand the detector response.
\red{CMOS sensors with smaller pixel sizes and thicker sensitive layers have been developed to meet requirements for higher timing resolution, angular resolution and efficiency compared to conventional CCD sensors onboard satellites \cite{arai11, tsuru18}}. Particularly for the hard X-ray band ($>$ 10 keV), the small pixel size enhances complex physical processes in sensors caused by incident photons including charge sharing among pixels, secondary electron ranges comparable to the pixel sizes, and contributions of various photon interactions \red{such as photoabsorption, Compton scattering, and fluorescence}.
\red{The charge sharing occurs when an X-ray photon is absorbed near the pixel edges. In this case, the energy of the incident photon and the interaction position can not be reconstructed without accurate knowledge of the charge cloud behavior.}
Therefore, a Monte Carlo simulation is necessary to understand the detector response to the hard X-ray \red{interactions}.

\begin{figure}[htb]
\begin{center}
\includegraphics[width=7cm]{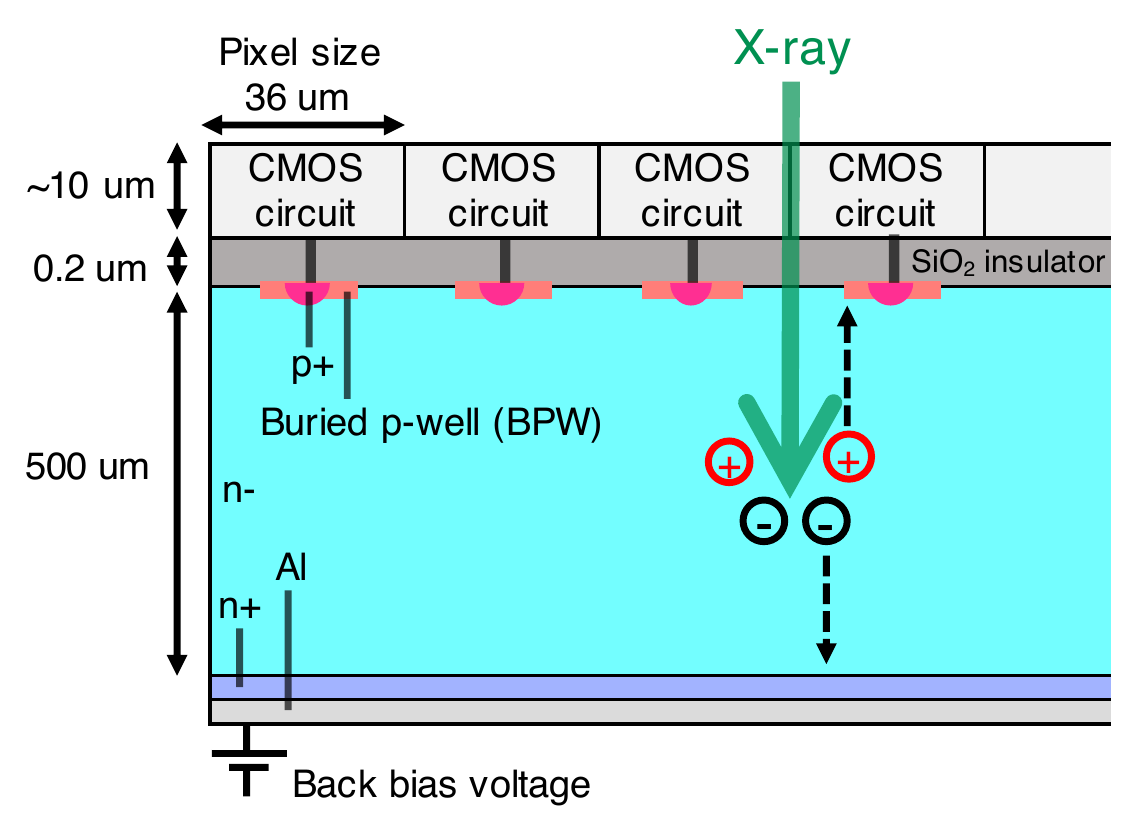}
\caption{Schematic drawing of a cross-sectional view of the XRPIX 6H sensor. \red{The carriers produced by an X-ray photon drift toward the sense nodes.} \label{fig-xrpix}}
\end{center}
\end{figure}

A proposed Japan--US collaboration satellite mission named FORCE (Focusing On Relativistic universe and Cosmic Evolution) is aimed at a wide-band imaging spectroscopy in 1--80 keV with an excellent angular resolution of $<$ 15 arcseconds as a half power diameter \cite{mori16, nakazawa18}. The main detector, Wideband Hybrid X-ray Imager (WHXI), is planned to be composed of two sensors: a silicon pixel sensor using Silicon-On-Insulator (SOI) technology which is called X-ray pixel (XRPIX) series \cite{arai11, tsuru18} and a cadmium telluride double-sided strip detector. \red{The requirements for the XRPIX sensor include a sensitive layer thicker than 200 $\mu$m, an energy resolution better than 300 eV in a full width at half maximum at 6 keV, and a timing resolution shorter than 15 $\mu$s.} The high timing resolution is required to \red{implement} an anti-coincidence background rejection technique using active shield counters.
Fig.~\ref{fig-xrpix} is a schematic cross-sectional view of the XRPIX 6H sensor. The XRPIX is a monolithic sensor of high resistivity Si on which CMOS circuits are implemented with SiO$_2$ insulator layer. The thick sensitive layer, the high-speed readout and the low readout noise can be realized by this SOI technology.

\red{The detector response of the XRPIX sensors has been investigated in several works.} 
\red{Matsumura\,et\,al.\,(2015)} \cite{matsumura15} and Negishi\,et\,al.\,(2019) \cite{negishi19} studied the spatial distribution of the charge collection efficiency (CCE) and an electric field structure in the sensors in order to evaluate the XRPIX sensor performance.
Hagino\,et\,al.\,(2019) \cite{hagino19} modeled the charge cloud size and verified it by comparing it to measured number fractions of charge-sharing events.
However, there \red{have} been no response modeling studies of the XRPIX sensors for hard X-ray \red{photon interactions} including charge-sharing event spectra, which are necessary for the detector designing and \red{evaluating the scientific capabilities} of FORCE.
In this work, we have developed a detector simulation framework for the XRPIX sensors, which can simulate the expected detector count spectrum from an arbitrary celestial source emission.

%%%%%%%%%%%%%%%%%%%%%%%%%%%%%%%%%%%%%%%%%%
\section{Design of the simulation framework}
\red{Our simulation framework is mainly composed of three parts: a Monte Carlo simulation of photon interactions based on Geant4 \cite{agost03, allison06, allison16}, a charge cloud transport simulation including the drift and diffusion of carriers, and an event extraction process.}
\red{All of these are implemented in ComptonSoft \citep{odaka10}}.
%\red{In our simulation, first the input emission spectrum and spatial distribution are modified assuming the telescope parameters and then the modified photon distribution is inputted to the sensor which is placed on the telescope's focal plane.}

\begin{figure*}[htb]
\begin{center}
\includegraphics[width=11cm]{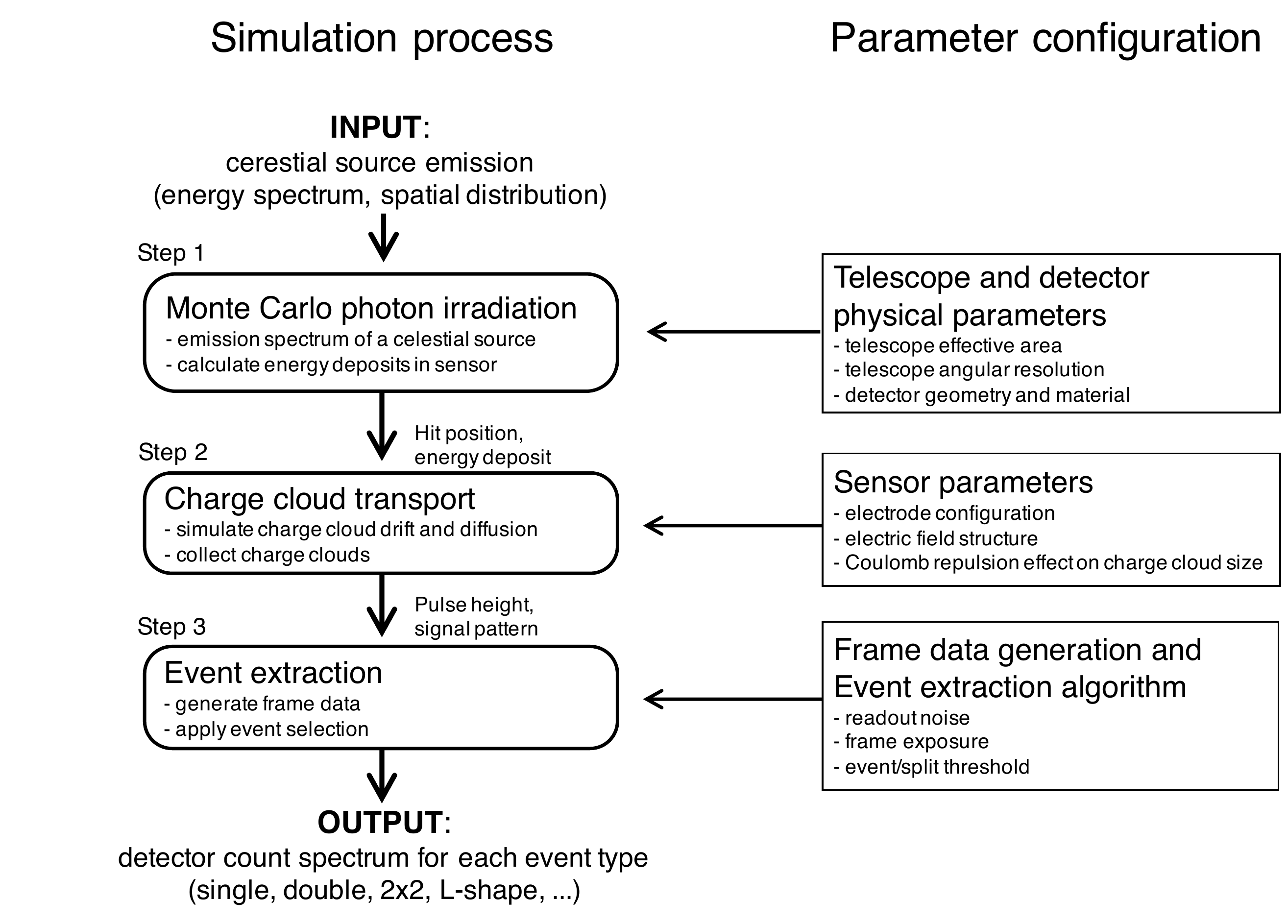}
\caption{Outline of our simulation procedures. \label{fig-sim}}
\end{center}
\end{figure*}

Fig.~\ref{fig-sim} shows the structure of the simulation framework.
First, in the Monte Carlo photon irradiation part (Step 1), we input the photon spectrum and the angular distribution of a celestial source. We input the effective area of an X-ray telescope as a function of energy and \red{the telescope's point spread function in order to get the photon spectrum and spatial distribution} at the time when these photons enter the sensor. \red{Then the Monte Carlo simulation is conducted based on Geant4, in which primary photons are generated just in front of the sensor and their interactions in the detector are simulated.} This part produces a list of interaction positions and energy deposits in the sensor.

In the charge cloud transport part (Step 2), for each interaction we simulate the drift and diffusion of the charge cloud, and calculate their collection efficiency. \red{Here we divide a charge cloud into a large number of point charges}, and calculate their transport by an input three-dimensional electric field structure and thermal diffusion plus a Coulomb repulsion effect.
\red{A total CCE for each interaction is calculated as the average over the CCEs for the divided point charges.
If the point charges are fully collected into sense nodes, which is a desired behavior, the CCE of this charge cloud is unity.
If a point charge is lost in the sensor, the CCE for this point charge is determined according to the weighting potential \cite{he01}.
However, since the exact form of the weighting potential is non-trivial due to the complicated structure of the pixel circuits, we assign the CCEs phenomenologically in this case (see Sec.~\ref{sec-comparison}).}
The Coulomb repulsion effect is treated phenomenologically by a magnification factor of the diffusion coefficient to the one which would be expected if only thermal diffusion were considered.
We calculate the thermal diffusion equation with a diffusion coefficient of $kT \mu_{\rm p}/e$, where $\mu_{\rm p}$ is the mobility of the holes ($5.0 \times 10^2$ cm$^2$ V$^{-1}$ s$^{-1}$). The drift time $t_{\rm d}$ is calculated as a numerical integration of the equation $dt_{\rm d} = |d{\bm x}|/|{\bm v}_{\rm p}({\bm x})|$, where ${\bm v}_{\rm p} ({\bm x})$ is the drift velocity of the holes. We consider the saturation of the drift velocity as $v_{\rm p} ({\bm x}) = v_{\rm s}/(1 + E_{0}/E({\bm x}))$ \citep{caughey67}. 
A saturation velocity $v_{\rm s}$ of $1.0 \times 10^7$ cm s$^{-1}$ and $E_{0}$ of $2.0 \times 10^4$ V cm$^{-1}$ are assumed. We also take into account the lifetime of the holes (10 $\mu$s is assumed) \cite{schroder97}.
As an output of this step, we generate a list of the pulse hight distribution for each photon interaction, which contains the pulse hight values induced in individual pixels.

The final step of our simulation is the event extraction part (Step 3).
We generate frame data with an assumed frame exposure time. In this process, an input readout noise is applied to each signal. Then an event selection algorithm with event/split threshold values as parameters is applied, and the events are classified into several types (e.g., single-/double-pixel events; see Fig.~\ref{fig-comp} for the definition of each event type).
The event threshold is a minimum signal value for each pixel to be judged as an X-ray event pixel.
The split threshold is a minimum signal value for each pixel to be judged as a charge-sharing pixel in the surrounding pixels of the central X-ray event pixel.

%%%%%%%%%%%%%%%%%%%%%%%%%%%%%%%%%%%%%%%%%%%%%%%
\section{Laboratory measurements}
We conducted laboratory measurements using an XRPIX sensor in order to make a comparison to the simulation results.
We used the XRPIX 6H sensor. It has a pixel size of 36 $\mu$m, a sensitive layer thickness of 500 $\mu$m, and is segmented into $48 \times 48$ pixels in total.
The sensor was operated in a thermostat chamber at $-40^\circ$C. A back-bias voltage of 300 V, which is larger than the value required for full depletion ($\sim$150 V), was applied \red{to suppress carrier diffusion and obtain high CCEs.}
The sensor was operated in the frame readout mode with a frame exposure of 1 ms.
We used radioactive isotopes (RIs) of $^{55}$Fe, $^{133}$Ba and $^{241}$Am in these measurements.

We applied a gain correction to each pixel. We derived the ADU-energy relation from the measured single-pixel event spectrum of each pixel and fitted it with a cubic function of energy, using the RI emission lines at 5.9 keV ($^{55}$Fe), 17.8 keV ($^{241}$Am) and 30.9 keV ($^{133}$Ba).
For the processing of the measured data, we used the same algorithm as the event extraction process in our simulation (Step 3 in Fig.~\ref{fig-sim}), and generated output data with the same format as the simulation results.
The event and split thresholds were set to be 1.5 keV and 0.45 keV ($\sim$10 and $\sim$3 times the standard deviation of the readout noise), respectively.

%%%%%%%%%%%%%%%%%%%%%%%%%%%%%%%%%%%%%%%%%%%%%%55
\section{Determination of the physical parameters of the XRPIX 6H sensor}\label{sec-comparison}
By comparing the simulation results to the measurements, we investigated whether our simulation with a reasonable parameter set of the XRPIX 6H sensor can reproduce the laboratory measurements.
The emission lines used for the comparison were the K$\alpha$ and K$\beta$ emissions from RIs of $^{55}$Fe ($\sim$5.9 keV) and $^{133}$Ba ($\sim$30.9 keV).
The geometry model of the simulation contained the XRPIX 6H sensor and its surrounding materials to describe the laboratory measurement environment.
The readout noise was fixed to 0.16 keV, which was determined from the measured single-pixel events' spectral line broadening at 5.9 keV.
In the event extraction part, we set the same event/split threshold values as those applied to the measurements.
The physical properties considered as the free parameters included the electric field structure and the Coulomb repulsion factor on the charge cloud size.
The Coulomb repulsion factor has been estimated analytically assuming a simple condition (spherically symmetric Gaussian distribution of a charge cloud) \cite{hagino19}, but this condition may not be applied to actual sensors.
Thus we treated this factor as a free parameter.

\begin{figure*}[h!]
\begin{center}
\includegraphics[width=10cm]{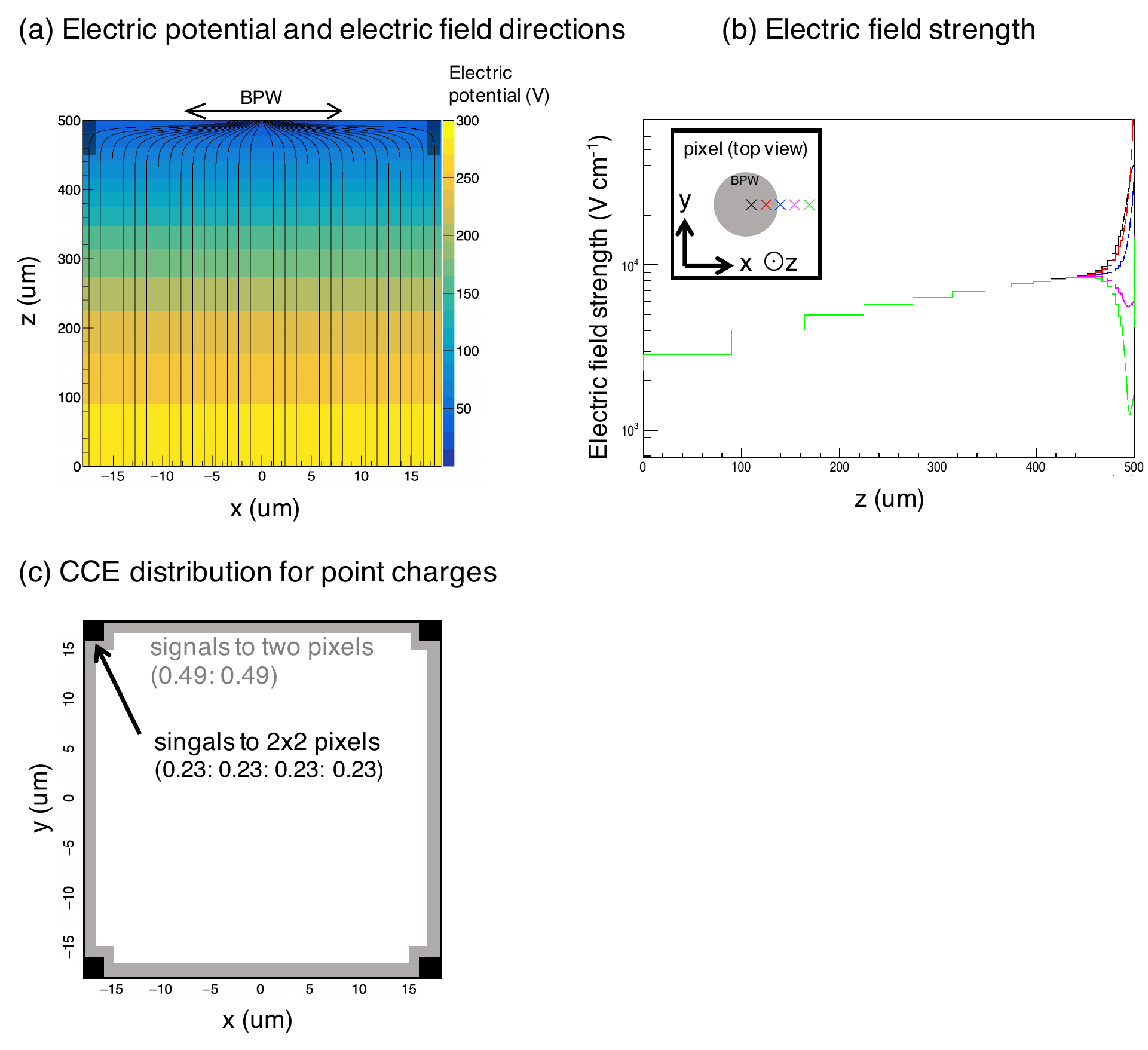}
\caption{The optimized physical parameter configuration of the XRPIX 6H sensor. (a) The colors and the solid lines represent electric potential values and the electric field directions, respectively.
Note that the electric field directions were obtained by test-particle tracking, and thus the field strength is not taken into account. The transparent black boxes at the pixel edges are the degraded regions (see text).
(b) The electric field strength as a function of the depth. The colors distinguish the (x,y) positions from which the electric field values were extracted \red{as shown in the top view of the pixel embedded in the plot. The electric field lines are concentrated toward the BPW region.
(c) The CCE distribution for point charges in the degraded region. The gray and black regions represent the area in which signals to two and 2x2 adjacent pixels are induced, respectively. The CCE fractions for its own and adjacent pixels are shown in parentheses.
}  \label{fig-param}}
\end{center}
\end{figure*}

In order to estimate the electric field structure in the sensor, we conducted a device simulation using \red{Hyper Device-Level Electrical Operation Simulator (HyDeLEOS), a part of the Technology Computer Aided Design (TCAD) system Hyper Environment for Exploration of Semiconductor Simulation (HyENEXSS) \cite{tcad}}. This simulator calculates the electric potential and the charge distribution in a semiconductor device. In the simulation, the implants and the substrate were implemented based on the parameters provided by Lapis Semiconductor Co. Ltd. According to the previous study, a positive fixed charge of $3\times 10^{11}$~cm$^{-2}$ was added at the insulator layer \cite{matsumura15}.
The electric potential and the electric field structure we obtained are shown in Fig.~\ref{fig-param} (a) and (b).
In Fig.~\ref{fig-param} (a), the cross-sectional view of the electric potential is shown with colors, and the electric field directions are drawn with solid black lines. Fig.~\ref{fig-param} (b) represents the electric field strength as a function of the depth. The colors indicate different extraction positions. The electric field strength becomes weaker toward the pixel edge.

\begin{figure*}[h!]
\begin{center}
\includegraphics[width=13cm]{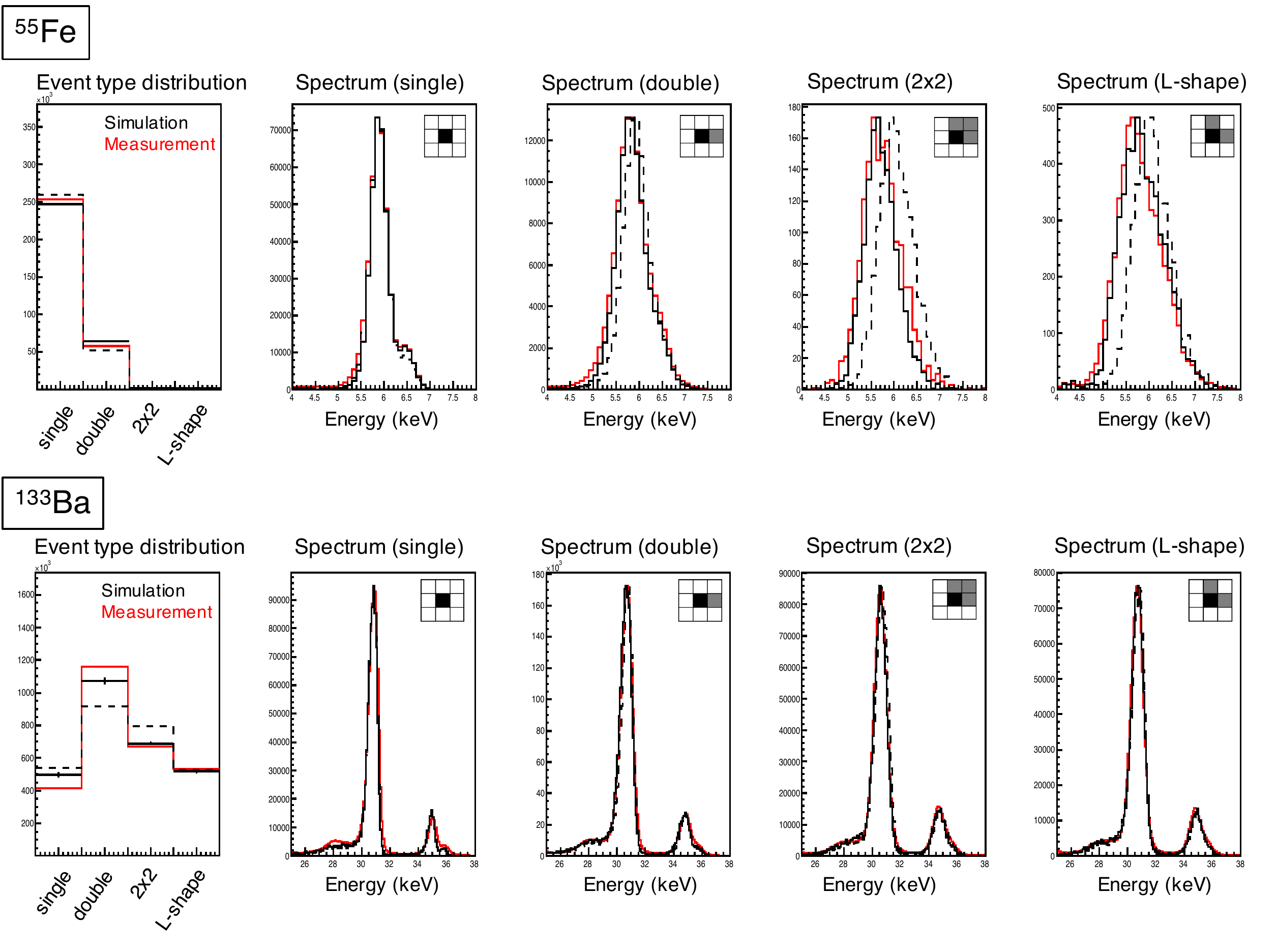}
\caption{Comparison between the simulations and the laboratory measurements. The number fraction and energy spectrum of each event type (single, double, 2x2 and L-shape) are shown. The red lines represents the measurement results. The dashed and solid black lines represent the simulation results before and after the optimization of the parameters, respectively. Note that, for comparison, \red{the event type distribution histograms are scaled so that the total event numbers} become the same for the three cases. The spectra were also scaled so that the peak values become the same, in order to compare the spectral shapes. The schematic pictures of 3x3 pixels shown in the panels represent shape definitions of the four event types. The black and gray pixels correspond to the X-ray event pixels and the charge-sharing pixels, respectively. \label{fig-comp}}
\end{center}
\end{figure*}

\begin{figure*}[h!]
\begin{center}
\includegraphics[width=10cm]{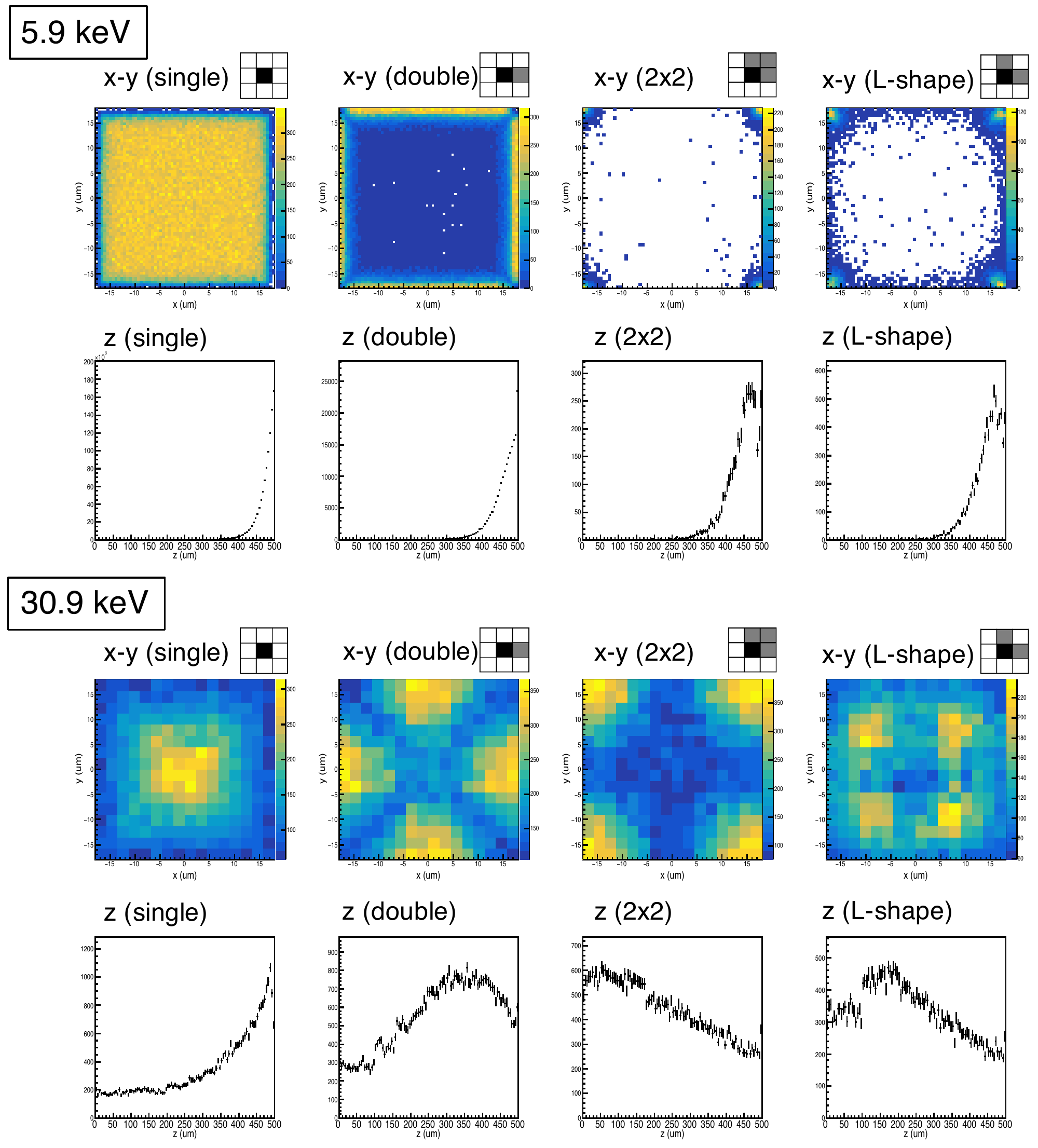}
\caption{\red{Spatial distribution of the interaction position within the X-ray event pixel for different event types. The 3x3-pixel drawings, which represent event types, are shown in the same way of Fig.~\ref{fig-comp}.} The interaction position distributions in the x-y plane (summed over the depth) and the vertical distributions (summed over the x-y plane) are shown for the 5.9 keV (upper panel) and the 30.9 keV (lower panel) irradiation. \label{fig-dist}}
\end{center}
\end{figure*}

First we assumed the sensor parameters as follows: the electric field structure which was obtained in the preceding TCAD simulation and the Coulomb repulsion factor of 1.45 at 5.9 keV and 1.55 at 30.9 keV, which were derived in the analytical calculations by Hagino\,et\,al.\,(2019) \cite{hagino19}.
Then the comparison showed that the simulation results had significantly higher line centroids in the charge-sharing event spectra ($^{55}$Fe case), and the lower number fraction of the charge-sharing events ($^{133}$Ba case) than the measurements had (see the dashed black lines and the red lines in Fig.~\ref{fig-comp}).
Considering the interaction position distributions of individual event types shown in Fig.~\ref{fig-dist}, these discrepancies were due to somewhat lower CCEs and stronger diffusion near the pixel edges, where most of the events were classified into the charge-sharing events.

Although the reason is yet to be \red{understood}, slight decreases of CCEs near the pixel edges in the XRPIX sensors have been reported \cite{matsumura14, negishi19, hagino19}. Also, the disruption and/or weakening of the electric field near the pixel edges, which \red{enhances} the charge cloud diffusion, \red{has} been discussed \cite{matsumura15}.
Thus, in order to fit the simulation to the measured data, we assumed that the carriers were lost near the pixel edges so that they make induced signals kept in both own and adjacent pixels, resulting in an increase of the charge-sharing event fraction.
Physically, this condition either corresponds to a significant decrease of the carrier velocities near the pixel edges because of high impurities, or to an insufficient electric-field line concentration toward the sense node near the pixel edges.
Both result in the interruption of the carrier transfer toward the sense node.

\red{In order to implement the effect considered above into our simulation model, we put degraded regions at the pixel edges (the black boxes in Fig.~\ref{fig-param} (a)). In the other regions, we kept the same electric field structure as above.}
Carriers that reach the degraded regions are immediately recombined and induce signals in their own and the adjacent pixels, \red{so that charge-sharing events are increased.}
\red{We assigned the CCEs for point charges lost in the degraded regions as shown in Fig.~\ref{fig-param} (c).
These CCEs for the lost charges in the degraded regions were set to be slightly less than unity: constant values of 0.98 near the pixel edges and 0.92 at the pixel corners (see Fig.~\ref{fig-param} (c)). This explains the slight line center shifts of the charge-sharing events toward lower energies.}
Finally, we modified the Coulomb repulsion factors to be 1.00 at 5.9 keV and 1.32 at 30.9 keV in order to fit the measured event type distribution.
Considering the thicker sensitive layer and the lower back-bias voltage than those in Hagino\,et\,al.\,(2019) \cite{hagino19}, the drift time ($t_{\rm d}$) of the carriers should be longer in our measurements.
Since the Coulomb repulsion factor is estimated to be roughly proportional to $t_{\rm d}^{-1/6}$ \cite{hagino19}, the factor is expected to be smaller in our measurements \red{as we obtained here, although it should be noted that this factor is introduced just for a phenomenological treatment}. 

The solid black lines in Fig.~\ref{fig-comp} represent the simulation results with the optimized parameters.
\red{Here the simulated spectral shapes of the charge-sharing events at 5.9 keV reproduce the measurements very well.}
\red{The simulated event type distributions well agreed with the measurements as well at both 5.9 keV and 30.9 keV within relative errors of $\sim10\%$ assuming realistic physical parameters of the sensor.
Although the residuals are not within statistical errors of the Monte Carlo simulations in Steps 1 and 2, this level of the discrepancy is acceptable considering our simple detector modeling.
These residuals are probably due to simplified assumptions of the electric field structure or the arrangement of the degraded regions, or the phenomenological treatment of the weighting potential.
These effects will be investigated with the future X-ray beam scanning.}

%%%%%%%%%%%%%%%%%%%%%%%%%%%%%%%%%%%%%%%%%%%%%%%%%%55
\section{Conclusion}
We have developed a simulation framework for the future X-ray satellite FORCE, which is a Japan-US collaboration mission aimed at an imaging spectroscopy in 1--80 keV.
In this framework, we can estimate the detector count spectrum from an assumed emission spectrum and angular distribution of a celestial source, by conducting a simulation on the detector response composed of photon \red{interactions}, charge cloud transport and an event extraction process.
As a validation of our simulations, using the XRPIX 6H sensor, we confirmed that the simulation results well agreed with our laboratory measurements assuming a physically reasonable parameter set of the sensor including the electric field structure, the Coulomb repulsion factor on the charge cloud size, and the arrangement of the degraded regions.
This simulation framework can be applied to the future updates of the XRPIX series as well as various types of semiconductor pixel sensors.

\section*{Acknowledgement}
We thank the anonymous referees for valuable advice, which improved the paper significantly.
We acknowledge the manufactures of XRPIXs and valuable advice by the personnel of LAPIS Semiconductor Co., Ltd.
This work was supported by the Japan Society for the Promotion of Science (JSPS) Research Fellowship for Young Scientist No. 19J11069 (H.S.), KAKENHI Grant-in-Aid for Scientific Research on Innovative Areas No. 25109004 (T.G.T.), Grant-in-Aid for Scientific Research Nos. 15H02090 (T.G.T.), 19H01906 and 19H05185 (H.O.), Challenging Exploratory Research No. 26610047 (T.G.T.), and Early-Career Scientists No. 19K14742 (A.T.).
This study was also supported by the VLSI Design and Education Center (VDEC), the University of Tokyo in collaboration with Cadence Design Systems, Inc., and Mentor Graphics, Inc.

\end{document}